\documentclass[10pt, conference]{IEEEtran}
\IEEEoverridecommandlockouts
\usepackage[left=0.60in, right=0.60in, bottom=0.98in, top=0.63in]{geometry} 
\usepackage{cite}
\usepackage{multicol}
\usepackage{amssymb}

\usepackage{amsmath}
\usepackage{amsfonts}
\usepackage[T1]{fontenc}
\usepackage{algpseudocode}
\usepackage[utf8]{inputenc}
\usepackage[english]{babel}
\usepackage[usenames, dvipsnames]{color}
\usepackage{pifont}
\usepackage{booktabs}
\usepackage{mathtools}
\usepackage{textcomp}
 \usepackage{graphicx}
\usepackage{subfigure}
\usepackage[ruled,vlined,commentsnumbered]{algorithm2e}
\usepackage{amsthm}
\usepackage{setspace}

\usepackage{multirow}

\SetKwInput{KwIn}{Inputs}

\theoremstyle{definition}

\theoremstyle{remark}

\theoremstyle{plain}

\newcommand{\RNum}[1]{\uppercase\expandafter{\romannumeral #1\relax}}

\def\BibTeX{{\rm B\kern-.05em{\sc i\kern-.025em b}\kern-.08em
    T\kern-.1667em\lower.7ex\hbox{E}\kern-.125em}}

\begin{document}
\title{Dynamic Interference Management for TN–NTN Coexistence in the Upper Mid-Band}
%Policy-Driven Coexistence of Terrestrial and Non-Terrestrial Networks in the Upper Mid-Band
% \author{\IEEEauthorblockN{ Pradyumna Kumar Bishoyi, Chia Chia Lee, Navid Keshtiarast, and Marina Petrova}
% 	\IEEEauthorblockA{Mobile Communications and Computing, RWTH Aachen University, Aachen, Germany \\
% 		Email: \{pradyumna.bishoyi, chia.lee, navid.keshtiarast, petrova\}@mcc.rwth-aachen.de}
% }
\author{
Pradyumna Kumar Bishoyi$^{\dagger}$, Chia Chia Lee$^{*}$, Navid Keshtiarast$^{*}$, and Marina Petrova$^{*}$\\[1pt]
$^{\dagger}$Department of Electrical Engineering, Indian Institute of Technology Jodhpur, Rajasthan, India\\
$^{*}$Chair of Mobile Communications and Computing (MCC), RWTH Aachen University, Aachen, Germany\\
Emails: pradyumna@iitj.ac.in, \{chia.lee, navid.keshtiarast, petrova\}@mcc.rwth-aachen.de
}

\maketitle

\begin{abstract}
The coexistence of terrestrial networks (TN) and non-terrestrial networks (NTN) in the frequency range 3 (FR3) upper mid-band presents considerable interference concerns, as dense TN deployments can severely degrade NTN downlink performance. Existing studies rely on interference-nulling beamforming, precoding, or exclusion zones that require accurate channel state information (CSI) and static coordination, making them unsuitable for dynamic NTN scenarios. To overcome these limitations, we develop an optimization framework that jointly controls TN downlink power, uplink power, and antenna downtilt to protect NTN links while preserving terrestrial performance. The resultant non-convex coupling between TN and NTN parameters is addressed by a Proximal Policy Optimization (PPO)–based reinforcement learning method that develops adaptive power and tilt control strategies. Simulation results demonstrate a reduction up to $8$ dB in the median interference-to-noise ratio (INR) while maintaining over $87\%$ TN basestation activity, outperforming conventional baseline methods and validating the feasibility of the proposed strategy for FR3 coexistence.
\end{abstract}

\IEEEpeerreviewmaketitle
%\vspace{-1em}
\section{Introduction}\label{sec:Introduction}
The rapid growth in global mobile data traffic, driven by widespread 5G adoption, immersive AR/VR applications, cloud gaming, and the increasing integration of artificial intelligence (AI), has created an urgent demand for additional spectrum resources in mobile systems \cite{TMelodia_mag, Minsoo_mag}. To address this, the International Telecommunication Union (ITU) at the World Radiocommunication Conference (WRC-23) agreed to study the upper mid-band frequency (7–24 GHz), also known as frequency range 3 (FR3), for potential International Mobile Telecommunications (IMT)  \cite{Hassan2023, Niloy2023}. However, most of the FR3 frequencies are already occupied by incumbent satellite and fixed-service systems, leading to inevitable spectrum overlap between emerging terrestrial 5G/6G deployments and existing non-terrestrial network (NTN) operations\cite{RKF2021,RSAreport}. This overlap raises critical concerns about mutual interference, particularly the interference from dense TN deployments toward the more interference-sensitive NTN downlink receivers \cite{SpaceX2022}. A proper interference characterization and management strategy within the FR3 band is, therefore, crucial to ensure NTN link protection while maintaining terrestrial network (TN) performance.

Recent studies by the satellite and mobile communication industries \cite{RKF2021, RSAreport, SpaceX2022} have investigated the feasibility of TN–NTN coexistence within the FR3 band. Reports in \cite{RKF2021,RSAreport} suggest that coexistence is generally feasible and predicted a minimal interference impact (<$1\%$ outage) on low earth orbit (LEO) satellite user terminals. Reports in \cite{SpaceX2022}, however, predict severe non-geostationary satellite orbit (NGSO) interference with outage probabilities exceeding $70\%$ in certain scenarios, underscoring the vulnerability of satellite receivers under dense TN deployments. These contrasting findings motivate further investigation into adaptive interference mitigation within the FR3 band. Existing works have therefore primarily focused on spatial suppression techniques such as interference-nulling beamforming \cite{icc_nulling2024}, advanced precoder design \cite{DCabric_dyspan_2025}, and dynamic exclusion-zone strategies \cite{ASCENT2024, Hassan2023, Niloy2023}. In \cite{icc_nulling2024}, an interference-nulling beamforming approach is proposed for the $12$~GHz band, where TN base stations (BSs) steer spatial nulls toward the satellite link to suppress TN-to-NTN interference. Similarly, \cite{DCabric_dyspan_2025} introduces a hybrid true-time-delay planar-array precoder for three-dimensional null steering. While these methods can effectively reduce cross-system interference, they require precise channel state information (CSI) between the TN and satellite, which is difficult to obtain in practice and highly sensitive to beam misalignment. On the other hand, dynamic exclusion-zone approaches \cite{ASCENT2024, Hassan2023, Niloy2023} mitigate interference by muting TN base stations near NTN receivers based on satellite trajectory, but this often results in unnecessary coverage gaps and reduced terrestrial throughput.

These limitations highlight the need for more practical and adaptive coexistence strategies. Antenna downtilt control offers an efficient way to reshape the TN radiation pattern and suppress upward interference without relying on complex CSI or precise beam nulling \cite{jeff_TWC2024}. With the recent introduction of direct satellite-to-cell services, LEO satellites now serve mobile user terminals directly \cite{GunedKurt_2023}. While protecting fixed satellite gateways is relatively straightforward, mobile NTN users are more vulnerable to interference from TN base stations due to their mobility and variable exposure to sidelobe emissions. Under such dynamic conditions, traditional beamforming and precoding techniques, designed for static NTN receivers, become insufficient. Hence, a joint optimization of TN antenna downtilt and transmit power is required to adaptively mitigate interference toward NTN users. Motivated by these challenges, in this work we develop a unified interference management framework for TN–NTN coexistence in the FR3 band. The contributions of this paper are summarized below.
%we propose a joint optimization framework that dynamically adjusts TN power, antenna downtilt, and muting decisions for dynamic interference management in TN–NTN coexistence within the FR3 band. 
\begin{itemize}
    \item We formulate an optimization framework that jointly captures the coupling between TN transmit power, antenna downtilt, and NTN interference constraints. This formulation explicitly models coexistence dynamics, where dense TN deployments must adapt their configurations to protect primary NTN downlink performance. 
    \item To address the non-convex and highly coupled nature of the formulated problem, we develop a Proximal Policy Optimization (PPO)–based reinforcement learning (RL) solution that dynamically learns optimal TN control policies. The proposed method adaptively tunes TN transmit power and antenna downtilt to mitigate NTN interference.
    %The proposed method adaptively adjusts TN transmit power and antenna downtilt in response to varying interference conditions, achieving an efficient trade-off between NTN protection and TN spectral efficiency.
    \item The simulation results validate the effectiveness of the proposed framework for both static and mobile NTN users. The results demonstrate interference suppression up to $6$-$8$ dB compared to conventional fixed-parameter and exclusion-zone strategies while maintaining over $ 87\%$ TN gNB availability without extensive muting.
    %, confirming the adaptability and robustness of the proposed power–tilt control scheme under realistic FR3 coexistence scenarios.
\end{itemize}

%These limitations highlight the need for more practical and adaptive coexistence strategies. Antenna downtilt control provides a simpler means of reshaping the TN radiation footprint and reducing upward interference without relying on complex CSI or beam nulling. Furthermore, combining downtilt control with transmit power adaptation and selective muting can achieve a more balanced trade-off between NTN protection and TN performance. Motivated by these insights, this work proposes a joint optimization framework that dynamically adjusts TN power, antenna downtilt, and muting decisions to minimize interference toward NTN users. The main contribution of work is as follows

%The usual approach to facilitating this coexistence is through spatial and/or spectral separation 

\section{System model}
We consider a spectrum-coexistence scenario in the FR3 upper-mid band, where a TN and an NTN operate in the same geographical region and share the same frequency resources. The TN consists of set of $ \mathcal{N}= \{1, 2, \ldots, N\} $ 5G gNodeBs (gNBs) serving a set of $\mathcal{U}_{TN}= \{1, 2, \ldots, U_{TN}\}$ terrestrial users, while the NTN consists of a single LEO satellite serving a set of $\mathcal{U}_{NTN}= \{1, 2, \ldots, U_{NTN}\}$ satellite user terminals. A total system bandwidth $B$ is shared across the region. The geographical area under study lies fully within the instantaneous LEO satellite footprint. 
%Both systems share the same total bandwidth $W$ in a region fully within the satellite footprint.

%Two categories of NTN terminals are considered: (i) fixed satellite ground stations equipped with dish-type user terminals, and (ii) Land Earth Stations in Motion (L-ESIM), representing mobile NTN receivers.
The primary objective of this work is to characterize and mitigate the interference generated by the TN downlink and uplink transmissions toward NTN downlink receivers. In the following subsections, we present the TN and NTN system parameters in depth, followed by the interference-to-noise ratio (INR) characterization. %and the NTN downlink throughput expression.
         
\subsection{NTN modelling}
In the NTN model, the link budget is primarily determined by the propagation geometry and the satellite antenna radiation characteristics. 
%Therefore, we first describe the NTN pathloss model, followed by the antenna model.
\subsubsection{NTN pathloss}
The NTN links are modeled according to 3GPP TR 38.811 \cite{Rel38_1}. The \textit{slant distance} between the satellite and the NTN user $d$ at elevation angle $\alpha$ is given by
\begin{equation}
    d=\sqrt{R_E^2\sin^2{\alpha}+h_0^2+2h_0R_E}-R_E\sin{\alpha},
\end{equation}
where $h_0$ is the satellite altitude and $R_E$ is Earth's radius. For a given $d(\alpha)$ and carrier frequency $f_c$, the free space path loss (FSPL) is expressed as \cite{Rel38_1},
\begin{equation} \label{eq:FSPL}
    PL_{FS}[dB] = 32.45 + 20\log_{10}(d) + 20\log_{10}(f_c),
\end{equation}
Beyond FSPL, the NTN link suffers from clutter loss ($L_{CL}$), atmospheric losses, and shadow fading ($L_{SF}$), both of which depend on whether the link is LOS or Non-Line-of-Sight (NLOS). Thus, the total NTN path loss ($PL_{NTN}$) is expressed as $PL_{NTN} = PL_{FS} + L_{CL} + L_{SF} + L_{Atm} + L_{IS} + L_{TS}$, 
% \begin{equation}\label{eq:PL_b}
%     PL_{NTN} = PL_{FS} + L_{CL} + L_{SF} + L_{Atm} + L_{IS} + L_{TS},
% \end{equation}
where $L_{Atm}$ is the atmospheric loss, $L_{IS}$ is the ionospheric scintillation loss, and $L_{TS}$ is the tropospheric scintillation loss, and the values are considered as specified in \cite{Rel38_1}.
% \begin{figure}[htbp]
%     \centering
%     \includegraphics[width=0.4\textwidth]{Figs/NTNPathLoss.png}
%     \caption{NTN Path Loss}
%     \label{fig:NTN Path Loss}
% \end{figure}
\subsubsection{NTN antenna model}
The LEO satellite is positioned at an altitude of $600$ km above Earth's surface and provides communication service to a fixed footprint region of \mbox{$20\times20$ $\text{km}^2$} through an always-on spot beam. We adopt the antenna model defined in 3GPP TR 38.811 \cite{Rel38_1}, and its gain pattern is expressed as
\begin{equation}\label{eq:ant_gain}
\begin{aligned}
G_{NTN} =
\begin{cases}
G_{NTN}^{\max}, & \text{for } \theta = 0, \\[4pt]
4\,G_{NTN}^{\max}\!\left|\dfrac{J_{1}(k a \sin\theta)}{k a \sin\theta}\right|^{2}, &
\text{for } 0 < |\theta| \le 90^{\circ}
\end{cases}
\end{aligned}
\end{equation}
where $G_{NTN}^{max}$ dBi is the peak antenna gain, $J_{1}(\cdot)$ is the Bessel function of the first kind, $a$ is the radius of the antenna's circular aperture, $k=2\pi f_c/c$ is the wave number, and $\theta$ is the angle measured from the bore sight of the antenna's main beam. 
%Note that, for our analysis, we fix the beam pointing and footprint to ensure stable, quasi-static NTN downlink coverage. $c$ is the speed of light in a vacuum, 
%Beam pointing and footprint are assumed fixed during the evaluation interval, representing quasi-static coverage for NTN downlink operation. 
%with peak antenna gain $G_{NTN}^{max} = 38.5$ dBi and EIRP density of $4$ dBW/MHz \cite[Table~6a.2.2.1-2]{Rel38_1}. 
%We consider a single LEO satellite located at an altitude of $600$ km above Earth's surface, with an always-on spot beam covering a footprint of $20\times20$ km². The main beam has an antenna gain $G_{NTN}^{max} = 38.5$ dBi and an EIRP density of $4$ dBW/MHz \cite[Table~6a.2.2.1-2]{Rel38_1}. 

% \begin{equation}
% \begin{split}
% &\text{max Tx power}[\text{dBm}] = \\ &\text{EIRP density} + 30 + 10 \log_{10}(N_{\text{RB}} \times \text{SCS} \times 12) - \text{MaxGain}
% \label{eq:satellite tx}
% \end{split}
% \end{equation}
\subsection{TN modelling}
Following the NTN link characterization, we now define the TN channel and antenna parameters. These parameters directly influence the interference power received by NTN terminals, as discussed later in the INR analysis.
\subsubsection{TN pathloss}
The large-scale propagation loss of TN link is modeled using the Urban Macro (UMa) pathloss model, as specified in 3GPP TR 38.901 \cite{Rel38_901}. The UMa model captures both LOS and NLOS propagation conditions. For the LOS case, the model defines two distance-dependent expressions based on whether the horizontal 2D distance $d_{2D}$ is smaller or larger than the breakpoint distance $d_{th}'$. The corresponding LOS path loss is given by
\begin{equation}\label{eq:UMa_LOS}
\begin{split}
&PL_{LOS}^{UMa} = \\[2pt]
&\begin{cases}
32.4 + 20\log_{10}(d_{3D}) + 20\log_{10}(f_c), & 10\ \mathrm{m} \le d_{2D} < d_{th}',\\[4pt]
\begin{aligned}[t]
&32.4 + 40\log_{10}(d_{3D}) + 20\log_{10}(f_c)\\[-0.2pt]
&\quad -\,10\log_{10}\big(d_{th}'^{2} + (h_{BS}-h_{UT})^{2}\big)
\end{aligned}
& d_{th}' \le d_{2D} \le 5\ \mathrm{km}.
\end{cases}
\end{split}
\end{equation}
where $d_{3D}$ is the 3-D distance. $h_{BS}$ and $h_{UT}$ represent the height of the gNB and the TN user terminal, respectively. 
%Further, the NLOS pathloss expression is,
% \begin{equation}\label{eq:UMa_NLOS}
% \begin{aligned}
% PL_{NLOS}^{UMa} &= 13.54 + 39.08\,\log_{10}(d_{3D}) + 20\,\log_{10}(f_c) \\[-2pt]
% &\quad - 0.6\,(h_{BS} - 1.5).
% \end{aligned}
% \end{equation}
\subsubsection{TN antenna modelling}
We model the radiation characteristics of TN gNB antennas in accordance with 3GPP TR~38.901~\cite{Rel38_901}. Each gNB antenna pattern is defined by three parameters: the horizontal angle relative to the antenna boresight $\phi'$, the vertical angle $\theta'$, and the downtilt angle $\theta_d$. For a given $\phi'$ and $\theta'$, the antenna gain of TN gNB is
\begin{equation}
G_{BS}(\theta',\phi') = G_{\max} - \min\!\left\{-\!\big[G_V(\theta') + G_H(\phi')\big],\, A_{\max} 
\right\}.
\end{equation}
where $G_{\max}$ is the maximum directional gain, and $A_{\max}$ is the maximum front-to-back attenuation limit. The horizontal and vertical gain are given by $G_H(\phi')=-\min\ \left[12\left(\frac{\phi'}{\phi'_{3\text{dB}}}\right)^{2},\, A_{\max}\right]$, and $G_V(\theta')=-\min\!\left[  12\left(\frac{\theta'- \theta_d}{\theta'_{3\text{dB}}}\right)^{2},\, SLA_V\right]$, respectively, 
% \begin{align}
% G_H(\phi') &= -\min\!\left[
%   12\left(\frac{\phi'}{\phi'_{3\text{dB}}}\right)^{2},\, A_{\max}
% \right], \\[3pt]
% G_V(\theta') &= -\min\!\left[
%   12\left(\frac{\theta'- \theta_d}{\theta'_{3\text{dB}}}\right)^{2},\, SLA_V
% \right].
% \end{align}
with $\phi'_{3\text{dB}}$ and $\theta'_{3\text{dB}}$ being the horizontal and vertical half-power bandwidth (HPBW). $SLA_V$ represents the side-lobe attenuation.
% \begin{equation}
% \begin{split}
% &G_{UE}(\theta, \phi) = \\ &\max\left(G_{max,UE} - \min\left(12\left(\frac{\theta}{65}\right)^2 + 12\left(\frac{\phi}{65}\right)^2, 25\right), 0\right)
% \label{eq:GUE}
% \end{split}
% \end{equation}

\subsection{Interference-to-Noise Ratio (INR)}
Based on the established TN and NTN link models, we now analyze the mutual interference characteristics arising from their coexistence in the same frequency band. Since both systems share the FR3 spectrum, TN transmissions can introduce significant interference to the NTN downlink, which is particularly sensitive due to the long propagation path and limited link margin \cite{TMelodia_mag}. To quantify this effect, we consider the interference-to-noise ratio (INR) as a key performance indicator representing the ratio of aggregated interference power to the receiver noise floor. For each considered gNB location and antenna configuration, we compute the interference at each NTN user terminal.

%We consider two distinct interference scenarios: (i) downlink interference from TN gNBs and (ii) uplink interference from TN users affecting the NTN downlink. Throughout the analysis, the TN system is assumed to operate in time-division duplexing (TDD) mode, where either the downlink or the uplink is active at any given time. 

%The interference impact on each NTN user terminal is quantified using the interference-to-noise ratio (INR), which compares the aggregate received interference power to the NTN receiver noise floor. Now, we calculate the aggregated INR at NTN user terminal due to both downlink and uplink TN operations.
\subsubsection{Interference due to downlink TN}
The interference received at NTN user terminal $n \in \mathcal{U}_{NTN}$ from the $i \in \mathcal{N}$ TN gNB is
\begin{equation}
    I_{n,i}^{DL} = P_i^{gNB} + G_{BS}(\theta',\phi') + G_{NTN}^{UT}((\theta'_{n,i})) - PL(d_{n,i}),
\end{equation}
where $P_i^{gNB}$ is the transmitting power of $i^{\text{th}}$ gNB, $G_{NTN}^{UT}$ is the NTN user antenna gain, $\theta'_{n,i}$ is the angle between the $n^{\text{th}}$ NTN user terminal's boresight and $i^{\text{th}}$ gNB's interference axis, $PL(\cdot)$ is the pathloss, and $d_{n,i}$ is the distance between $n^{\text{th}}$ NTN user terminal and $i^{\text{th}}$ gNB. 

\subsubsection{Interference due to uplink TN}
Similarly, the interference received at NTN user terminal $n \in \mathcal{U}_{NTN}$ from TN user $i \in \mathcal{U}_{TN}$ is given by, $    I_{n,i}^{UL} = P_i^{UE} + G_{UE}(\theta',\phi') + G_{NTN}^{UT}((\theta''_{n,i})) - PL(d'_{n,i})$, 
% \begin{equation}
%     I_{n,i}^{UL} = P_i^{UE} + G_{UE}(\theta',\phi') + G_{NTN}^{UT}((\theta''_{n,i})) - PL(d'_{n,i}),
% \end{equation}
where $P_i^{UE}$ is the TN user uplink power, $G_{UE}$ is the user antenna gain, $\theta''{n,i}$ is the angular offset between the TN UE and NTN terminal, and $d'_{n,i}$ is the distance between TN and NTN user terminal.

\subsubsection{Aggregated INR}
Considering the effect of interference from both downlink and uplink TN transmission, the aggregated INR (in dB) at the NTN user $n \in \mathcal{U}_{NTN}$ is expressed as
\begin{equation}\label{eq:INR}
\begin{aligned}
INR_{n} &= 10\log_{10}\!\bigg( \sum_{i \in \mathcal{N}} I_{n,i}^{DL}\bigg) + 10\log_{10}\!\bigg( \sum_{i \in \mathcal{U}_{NTN}} I_{n,i}^{UL}\bigg)\\[-2pt] & - 10\log_{10}(kTB)
\end{aligned}
\end{equation}
% \begin{equation}\label{eq:INR}
% \begin{aligned}
% INR_{n} &= \mathbf{1}_{(\chi = 0)}\bigg[10\log_{10}\!\bigg( \sum_{i \in \mathcal{N}} I_{n,i}\bigg)\bigg] \\[-1pt] &+ \mathbf{1}_{(\chi = 1)}\bigg[10\log_{10}\!\bigg( \sum_{i \in \mathcal{U}_{NTN}} I_{n,i}^{UL}\bigg)\bigg] - 10\log_{10}(kTB)
% \end{aligned}
% \end{equation}
%$\chi \in \{0,1\}$ is a binary variable, such that $\chi = 0$ and $\chi = 1$ represent the downlink and uplink mode of TN system, respectively. Further, 
where $k$ represents the Boltzmann constant, $T$ is the temperature, and $B$ is the NTN user terminal bandwidth. Typically, INR quantifies the interference penalty experienced by an NTN user terminal \cite{Niloy2023}. An NTN terminal is considered interfered with when its INR exceeds a specified protection threshold. Typically, the protection threshold range spanning from $-12.2$ dB to $-6$ dB \cite{SpaceX2022, ASCENT2024}. 
\subsection{NTN aggregated downlink throughput}
The quantified INR directly impacts the achievable NTN downlink rate. The aggregated downlink throughput of the NTN system is determined by Shannon's capacity equation as follows,
\begin{equation}
    \mathcal{R}_{NTN} = B\sum_{n \in \mathcal{U}_{NTN}}\log_2(1+\gamma_n),
\end{equation}
where B represents the total NTN bandwidth. $\gamma_n$ represents the corresponding signal-to-interference-and-noise-ratio (SINR) of NTN user $n \in \mathcal{U}_{NTN}$, defined as  
\begin{equation}
    \gamma_n = \frac{P_{n}^{rx}}{\sum_{i \in \mathcal{N}} I_{n,i}+ \sum_{i \in \mathcal{U}_{NTN}} I_{n,i}^{UL} + N_0},
\end{equation}
where $P_{n}^{rx}$ is the received power at NTN user terminal and $N_0$ is the noise power spectral density.

The NTN downlink throughput depends on both TN transmit power and antenna configurations, which jointly determine the interference experienced by NTN users. We next formulate an optimization problem to identify optimal TN control parameters that preserve NTN performance while satisfying FR3 coexistence constraints.

\section{Problem formulation}
Building on the INR and throughput relationships, in this section, we formalize the optimization problem for TN–NTN coexistence. Interference from TN downlink transmissions poses a critical challenge to maintaining reliable NTN downlink performance. On the downlink side, adaptive transmit power control at TN gNBs can directly control the radiated energy toward the satellite footprint, while antenna downtilt adjustment provides an additional degree of freedom to reshape the vertical radiation pattern and suppress undesired emissions toward NTN users. Consequently, a coordinated control strategy that jointly optimizes gNB transmit power and antenna downtilt is essential to balance NTN protection and TN performance. 
%On the uplink side, transmit power control at TN user terminals becomes crucial to mitigate aggregated interference received at the NTN downlink, especially when a large number of TN user devices are active simultaneously. However, excessive power reduction in either direction can degrade TN service quality and coverage. Consequently, a coordinated control strategy that jointly optimizes gNB transmit power, antenna downtilt, and TN uplink transmit power is essential to balance NTN protection and TN performance. 
The optimization problem is formulated as
\begin{align}
\max_{\mathbf{P}^{gNB},\boldsymbol{\theta}_d} \quad 
& \mathcal{R}_{NTN}(\mathbf{P}^{gNB}, \boldsymbol{\theta}_d)
\label{eq:opt_obj}\\[4pt]
\text{s.t.} \quad
& INR_n(\mathbf{P}^{gNB}, \boldsymbol{\theta}_d) 
\leq \Gamma_{\text{th}}, \quad \forall n,
\label{eq:opt_inr}\\[6pt]
& \theta_{d,i}^{\min} \leq \theta_{d,i} \leq \theta_{d,i}^{\max}, \quad \forall i \in \mathcal{N},
\label{eq:opt_tilt}\\[6pt]
% & P_{\min}^{UE} \leq p_{j}^{UE} \leq P_{\max}^{UE}, \quad \forall j \in \mathcal{U}_{TN},
% \label{eq:opt_ulpower}\\[6pt]
& P_i^{gNB} \leq P_{\max}, \quad \forall i \in \mathcal{N}, \label{eq:opt_dlpower}
\end{align}
where $\mathbf{P}^{gNB} = [P_1^{gNB}, \dots, P_N^{gNB}]$ represents the transmit power vector of TN gNBs and $\boldsymbol{\theta}_d = [\theta_{d,1}, \dots, \theta_{d,N}]$ denotes the corresponding antenna downtilt configuration. Constraint~\eqref{eq:opt_inr} ensures that the INR experienced by each NTN user remains below the protection threshold $\Gamma_{\text{th}}$. The downtilt bound in~\eqref{eq:opt_tilt} limits the gNB tilt angle within its feasible mechanical range to control interference without compromising terrestrial coverage.  Finally, the downlink power constraint in~\eqref{eq:opt_dlpower} restricts each gNB’s transmit power below $P_{\max}$.
%Through~\eqref{eq:opt_ulpower}, we limit each TN user's uplink transmit power to the standard-compliant range 
%$[P_{\min}^{UE}, P_{\max}^{UE}]$ to sustain uplink connectivity.
%$\mathbf{P}^{UE} = [P_1^{UE}, \dots, P_{U_{TN}}^{UE}]$ denote TN uplink transmit power vector, 

The formulated optimization problem is inherently difficult to solve due to strong nonlinear coupling among TN transmit power ($\mathbf{P}^{gNB}$), and antenna downtilt ($\boldsymbol{\theta}_d$). Furthermore, the SINR and INR expressions involve logarithmic and fractional terms, rendering the objective function highly nonconvex and non-differentiable. Conventional convex optimization methods fail to capture these complex dynamics or guarantee convergence to a global solution. To overcome these limitations, we adopt a reinforcement learning (RL)-based approach to iteratively learn adaptive power–tilt control policies under dynamic interference environments.
%using the Proximal Policy Optimization (PPO) algorithm, which uses policy-gradient updates to iteratively learn adaptive power–tilt control policies under dynamic interference environments.
% The problem in \eqref{eq:obj} defines a joint power--tilt optimization framework for TN adaptation under TN--NTN coexistence. 

% \begin{equation}
% \begin{split}
%     \mathcal{O} &= \max \big[ \mathbb{Reward} \big] \\
%     &= \max \big[ 10 \cdot \text{NTN}_{\text{AvgThpt}} + 1 \cdot \text{TN}_{\text{ActivePct}} - 10 \cdot \text{NTN}_{\text{HarmfulPct}} \big] \\
%     &= \max \left[ 10 \cdot \underbrace{\left( \frac{1}{N_{\text{NTN}}} \sum_{i=1}^{N_{\text{NTN}}} \text{Thpt}_{i}^{\text{NTN}} \right)}_{\text{NTN Average Throughput}} + 1 \cdot \underbrace{\left( \frac{N_{\text{Active}}^{\text{TN}}}{N_{\text{Total}}^{\text{TN}}} \right)}_{\text{TN Active Percentage}} - 10 \cdot \underbrace{\left( \frac{N_{\text{Harmful}}^{\text{NTN}}}{N_{\text{Total}}^{\text{NTN}}} \right)}_{\text{NTN Harmful Percentage}} \right]
%     \label{eq:objective}
% \end{split}
% \end{equation}
% The first term maximizes the average NTN throughput, the second preserves TN activeness, and the third penalizes NTN terminals exceeding the interference threshold. The weighting ratio $10 : 1 : –10$ is determined empirically to prioritize protection and efficiency balance.
% The objective formulation defines the mathematical foundation of the coexistence control problem but does not assume any specific implementation method. The following section details the centralized PPO-based solution designed to realize this optimization.

\section{Proposed Solution Approach}
\begin{figure}[htbp]
    \centering
    \includegraphics[width=0.37\textwidth]{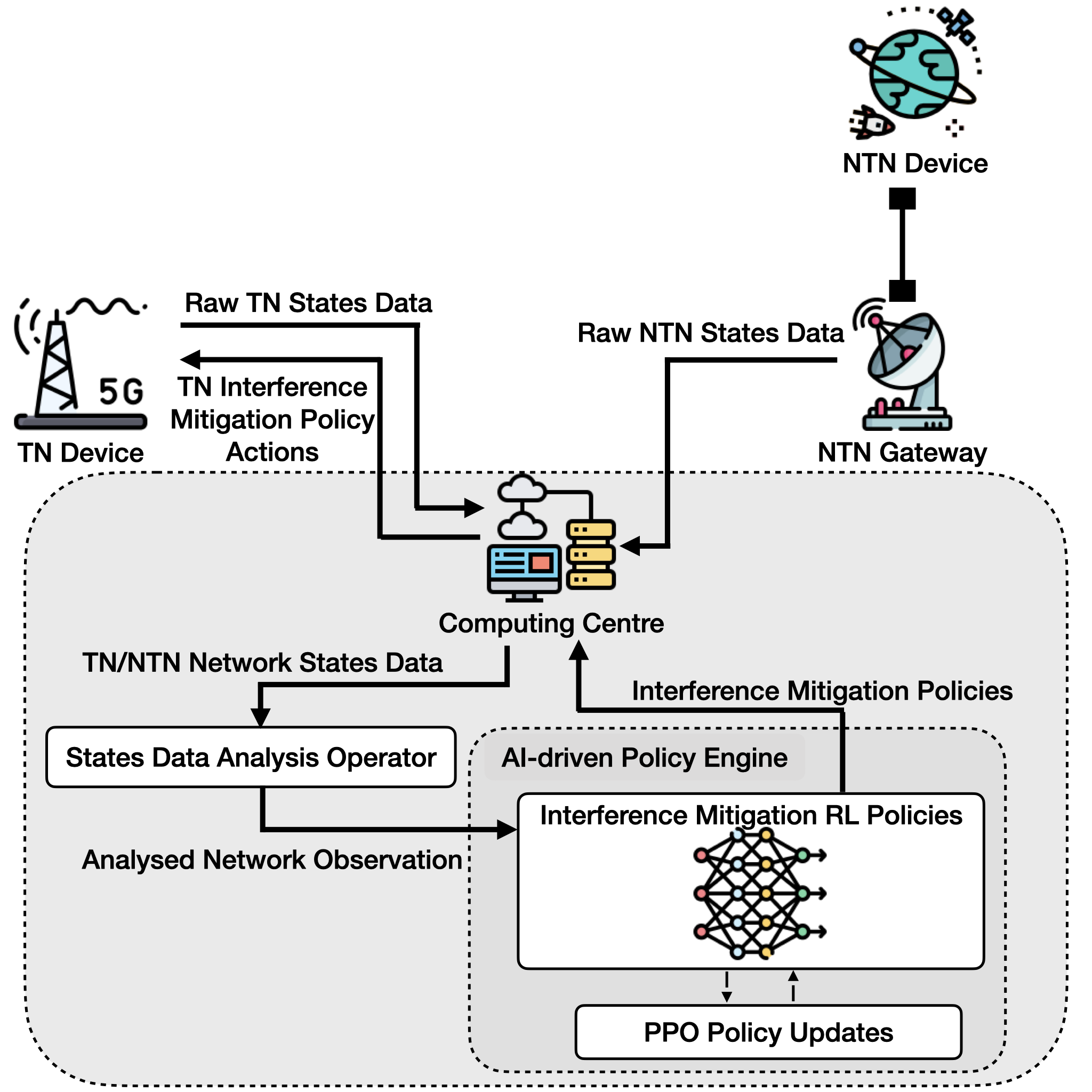}
    \caption{Interference Mitigation RL System}
    \label{fig:M4 Interference Mitigation RL system}
\end{figure}

To this end, we propose a centralized RL–based framework employing the Proximal Policy Optimization (PPO) algorithm. The centralized agent is deployed at the core network level, coordinating the joint adaptation of TN base station transmit power, user uplink power, and antenna downtilt across the shared FR3 spectrum. %Through continuous interaction with the environment, the PPO agent learns adaptive interference mitigation strategies that preserve NTN downlink performance while maintaining TN spectral efficiency.
The overall framework illustrated in Fig.~\ref{fig:M4 Interference Mitigation RL system}, operates through a closed-loop interaction between the TN-NTN systems and the centralized learning process. The TN and NTN entities periodically report network state information, such as interference levels, active gNB status, and NTN throughput, to the core network. The centralized PPO agent processes this information, evaluates interference conditions, and issues optimized control commands to the TN system. In addition to power and antenna downtilt control actions, we introduce a selective sector muting mechanism as a discrete control action. In dense urban deployments, certain BS sectors may continue to generate harmful interference toward NTN users even after power and tilt adjustments. Temporarily muting such sectors during critical NTN downlink periods provides a flexible and localized interference mitigation measure without requiring a full network shutdown. 
%We next define the observation and action space used by the PPO agent to model network states and determine optimal interference mitigation actions.
%At each discrete elevation step, the environment module provides the system observation state.  This state is passed to the PPO-based RL agent, which outputs an action vector containing sector muting, BS/UE power adjustment, BS tilting angle tuning, and threshold reconfiguration commands. The interaction loop evolves shown in Fig \ref{fig:Interference Mitigation Agent Training States} and work as follows:

\subsection{Observation and Action Space}
At each decision step $t$, the PPO agent observes a state vector $\mathbf{s}_t$ representing the interference and activity conditions across the TN–NTN coexistence environment, and selects an action vector $\mathbf{a}_t$ that jointly controls TN transmission and configuration parameters. The observation space $\mathcal{S}$ is defined as
\begin{equation}
\mathcal{S} = \left\{\mathbf{s}_t = \big[\theta_{EL,t}, \eta_t, \mathcal{R}_{NTN,t}, \chi_t, \Gamma_{th}\big]\right\},
\end{equation}
where $\theta_{EL}$ denotes the satellite elevation angle, $\eta_t$ represents the percentage of NTN user terminals experiencing harmful interference (i.e., $\text{INR} > \Gamma_t$), $\mathcal{R}_{NTN,t}$ is the aggregated NTN throughput, $\chi_t$ is the percentage of active TN gNBs, and $\Gamma_{th}$ represents the current INR protection threshold in use.

The action space $\mathcal{A}$ defines the control variables that the PPO agent can adjust at each step, and is given by
\begin{equation}
\mathcal{A} = \left\{\mathbf{a}_t = \big[\Gamma_{th,t}^{new},\delta_{i,t},P_{i,t}^{gNB},
P_{i,t}^{UE},\theta_{d,i,t}\big]\right\}, \quad i \in \mathcal{N},
\end{equation}
where $\Gamma_{th,t}^{new}$ represents the updated INR protection threshold; $\delta_{i,t} \in \{0,1\}$ denotes the binary sector muting control; $P_{i,t}^{gNB} \in [P_{\min}, P_{\max}]~\text{dBm}$ and $P_{i,t}^{UE} \in [P_{\min}^{UE}, P_{\max}^{UE}]~\text{dBm}$ are the downlink and uplink transmit powers of TN gNB and UEs, respectively, and $\theta_{d,i,t} \in [\theta_{d,i}^{\min}, \theta_{d,i}^{\max}]$ specifies the TN gNB antenna downtilt angle. 
% where $\Gamma_t^{new} \in [-12, -6]~\text{dB}$ is the updated INR protection threshold, 
% $\delta_{i,t}^{(on/off)} \in \{0,1\}$ represents the sector activation control,
% $P_{i,t}^{BS} \in [10, 33]~\text{dBm}$ and 
% $P_{i,t}^{UE} \in [10, 23]~\text{dBm}$ denote the TN BS and UE transmit powers, respectively, and 
% $\theta_{i,t}^{BS} \in [2^{\circ}, 15^{\circ}]$ is the BS antenna downtilt angle.  

\subsection{Reward function}
The immediate reward $r_t$ at time step $t$ is designed to guide the PPO agent toward maximizing NTN performance while ensuring TN continuity and limiting harmful interference. It is expressed as a weighted sum of normalized throughput, TN activity, and interference penalty terms, as expressed below
\begin{equation}\label{eq:reward_total}
r_t = w_1 \, \bar{\mathcal{R}}_{NTN,t} + w_2\,\chi_t - w_3\,\eta_t, 
\end{equation}
where $\bar{\mathcal{R}}_{NTN,t} = \frac{\mathcal{R}_{NTN,t}}{\mathcal{R}_{NTN}^{\max}}$ denotes the normalized NTN downlink throughput, and $\chi_t = \frac{N_{Active}}{N} $ denotes the fraction of active TN gNBs. The coefficients $w_1$, $w_2$, and $w_3$ control the trade-off between NTN throughput maximization, interference suppression, and TN service continuity. 
%This formulation ensures that the agent is rewarded for improving NTN throughput while maintaining adequate TN activity and penalizing excessive interference toward NTN users.

%The centralized agent learns an optimal policy $\pi(\mathbf{a}_t|\mathbf{s}_t)$ that maps the observed state $\mathbf{s}_t \in \mathcal{S}$ to an action $\mathbf{a}_t \in \mathcal{A}$ to maximize long-term NTN throughput.

% \begin{figure}[htbp]
%     \centering
%     \includegraphics[width=0.45\textwidth]{Figs/M4_PPO_training_agent.png}
%     \caption{Interference Mitigation Agent Training States}
%     \label{fig:Interference Mitigation Agent Training States}
% \end{figure}

%The PPO agent updates its policy parameters using the clipped-surrogate objective, ensuring learning stability under continuous, high-variance interference patterns. The training hyperparameters, covering discount factor, batch size, entropy regularization, and learning rates etc, is summarized in Table \ref{tab:hyperparameters}.

\subsection{PPO for TN-NTN coexistence}
We use the Proximal Policy Optimization (PPO) algorithm as the learning backbone of the proposed centralized framework, balancing stable policy improvement with efficient training. PPO is an actor–critic–based reinforcement learning method that maintains two neural networks \cite{Recmac,PPOAlgorithm}. An actor that parameterizes the policy $\pi_{\theta}(\mathbf{a}_t|\mathbf{s}_t)$ to select actions, and a critic that estimates the state-value function $V_{\phi}(\mathbf{s}_t)$ to evaluate the long-term quality of those actions. During training, the actor interacts with the TN–NTN environment to collect trajectories of states, actions, and rewards. The advantage function, $A_t = \hat{R}_t - V_{\phi}(\mathbf{s}_t)$, measures how much better an action performed is compared to the expected baseline value, and guides the policy update. To ensure stable learning, PPO maximizes a clipped surrogate objective that constrains large policy updates through a ratio term $r_t(\theta) = \frac{\pi_{\theta}(\mathbf{a}_t|\mathbf{s}_t)}{\pi_{\theta_{\text{old}}}(\mathbf{a}_t|\mathbf{s}_t)}$. The objective is given by $L^{\text{PPO}}(\theta) = \mathbb{E}_t \left[\min\left(r_t(\theta) A_t, \, \text{clip}\big(r_t(\theta), 1-\epsilon, 1+\epsilon\big) A_t\right)\right]$, 
% \begin{equation}
% L^{\text{PPO}}(\theta) = \mathbb{E}_t \left[\min\left(r_t(\theta) A_t, \, \text{clip}\big(r_t(\theta), 1-\epsilon, 1+\epsilon\big) A_t\right)\right],\nonumber
% \end{equation}
where $\epsilon$ is a clipping parameter that limits the deviation between new and old policies, preventing performance collapse due to overly aggressive updates. The critic network minimizes the value loss $\mathcal{L}_V = (V_{\phi}(\mathbf{s}_t) - \hat{R}_t)^2$, enabling better estimation of expected returns. 

Through alternating actor–critic updates, PPO efficiently learns stable and adaptive control strategies for power adjustment, antenna downtilt tuning, and sector muting. The following section presents the simulation setup and performance evaluation results to validate the effectiveness of the proposed learning-based coexistence framework.

% \begin{figure}[htbp]
%     \centering
%     \includegraphics[width=0.35\textwidth]{Figs/Satellite_orbit_map_1(NASA).png}
%     \caption{Satellite Constellation Setup Global Orbit}
%     \label{fig:Satellite Constellation Setup Global Orbit}
% \end{figure}
\begin{figure}[t]
    \centering
    \includegraphics[width=0.4\textwidth]{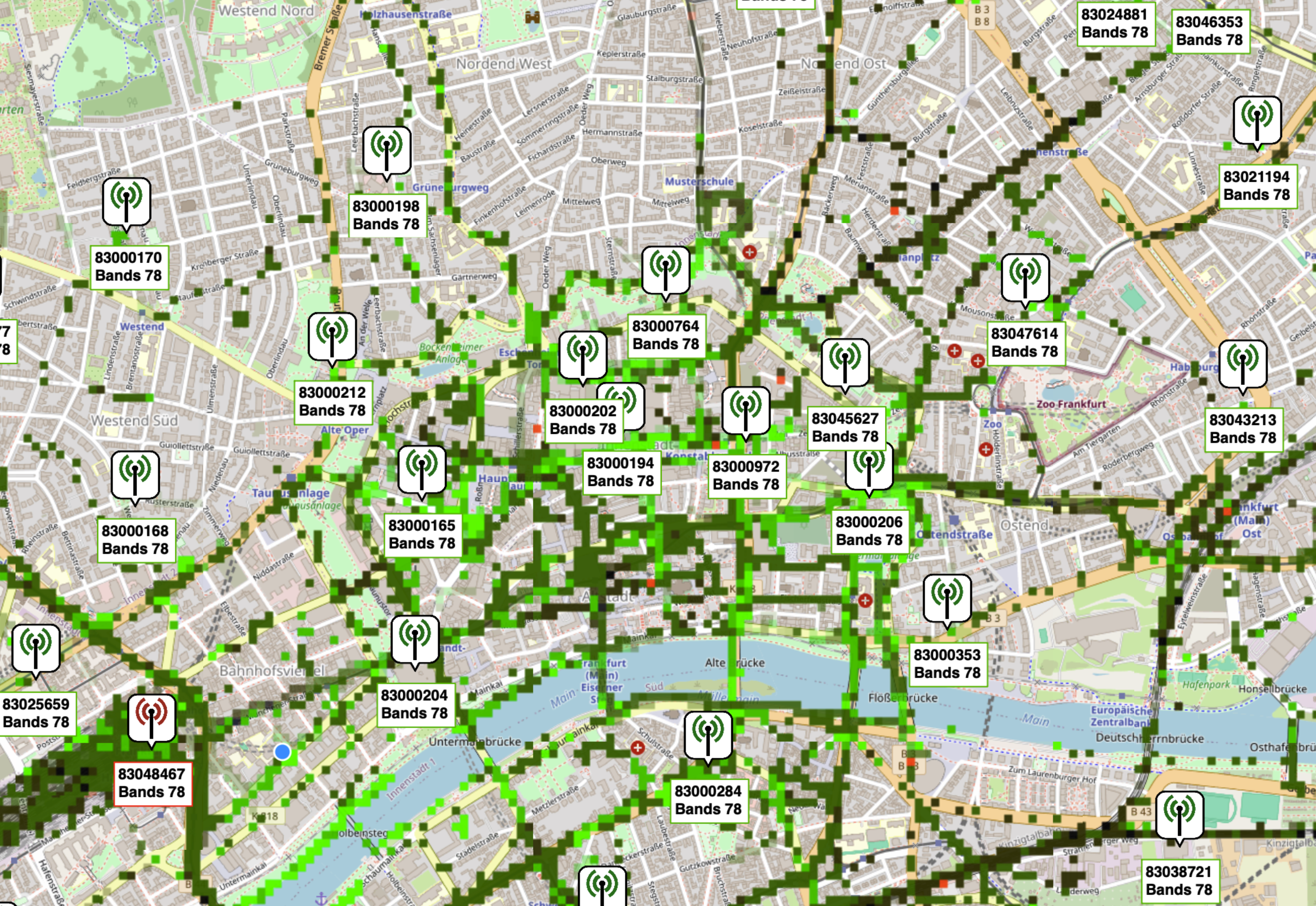}
    \caption{Base Station Map (Frankfurt)}
    \label{fig:Base Station Map}
\end{figure}

\section{Perfromance Evaluation}\label{sec:usecasesection}
%%%%%%%Navid's version
\subsection{Simulation setup}
We consider a TN-NTN coexistence scenario where NTN user terminals are spatially distributed according to a homogeneous Poisson Point Process (PPP) within a \(20\times20\) km\(^2\) LEO satellite spot-beam footprint. Two types of NTN users are considered: fixed gateway terminals (T1) with high-gain, narrow-beam antennas, and direct-satellite-to-cell mobile terminals (T2) with low-gain, wide-beam antennas. Simulations are performed for varying NTN user densities: \(6\times10^{-8}\), \(1\times10^{-7}\), and \(3\times10^{-7}\) users/m\(^2\), corresponding to 24, 40, and 120 active terminals, respectively. Additionally, different T1:T2 user ratios are examined to capture the impact of antenna characteristics and spatial load on TN-to-NTN interference. To capture dynamic link evolution during a satellite pass, one snapshot is recorded for every \(1^\circ\) change in the LEO satellite elevation angle. For a \(500\) km LEO orbit with \(53^\circ\) inclination, the snapshot interval varies from roughly $1$ s near zenith to $10$ s near the horizon. The full visible pass is defined over the \(10^\circ\) to \(170^\circ\) elevation, yielding $160$ simulation snapshots.

The terrestrial network layout is constructed from real base station data for Frankfurt am Main, mapped within the same $20\times20$ km² region, as shown in Fig. \ref{fig:Base Station Map}. Each BS retains its real-world coordinates, while TN user terminals are generated following a homogeneous PPP with density \(1.3 \times 10^{-5}\) users/m\(^2\) and associated with the nearest BS. All the key simulation parameters are summarized in Table \ref{tab:setup_parameters}.

% \begin{figure}[htbp]
%     \centering
%     \includegraphics[width=0.4\textwidth]{Figs/Simulation_Environment_Structure_3.png}
%     \caption{Simulation scenario.}
%     \label{fig:Simulation Environment Structure 3}
% \end{figure}

%\subsection{Baseline methodologies}
%%%%%%%%%%%Navid Version

\textbf{\textit{Baseline methodologies}:} To benchmark the effectiveness of the proposed PPO-based framework, we compare its performance against two baseline interference-management schemes. The first is a \textit{no-coordination} scenario, where the TN operates without any NTN-aware coordination or interference control, and the gNBs transmit with fixed power and antenna settings. The second is the rule-based \textit{ASCENT approach}~\cite{ASCENT2024}, which employs deterministic exclusion zones and threshold-based mute/unmute rules to protect NTN receivers, typically improving NTN protection at the expense of TN coverage.

\begin{table}[h!]
    \centering
    \caption{Environment setup parameters}
    \label{tab:setup_parameters}
    \begin{tabular}{|p{0.27\textwidth}|p{0.1\textwidth}|}
        \hline
        \textbf{Parrameters} & \textbf{Value} \\
        \hline
        Carrier frequency & 12 GHz \\
        \hline
        Total bandwidth & 200 MHz \\
        \hline
        % Number of LEO beams & 1 \\
        % \hline
        % Size of LEO Beams & $20\times20$ km² \\
        % \hline
       % Satellite Users per Beam & Low/Medium/High Density \\
        %\hline
        Number of gNBs in beam footprint & $106$ \\
        \hline
        UEs per TN gNB Cell & $12-20$ \\
        \hline
        Satellite Tx power & $48$ dBm \\
        \hline
        Satellite antenna gain & $38.3$ dBi \\
        \hline
        Dish terminal antenna gain & $33$ dBi \\
        \hline
        Direct-to-cell device antenna gain& $17$ dBi \\
        \hline
        % TN gNBs max antenna gain & $30$ dBi \\
        % \hline
        TN gNBs max Tx power  & $33$ dBm \\
        \hline
        % TN UEs max Tx gain & $11.0205$ dBi \\
        % \hline
        TN UEs max Tx power & $23$ dBm \\
        \hline
       % Satellite User Effective Temperature & $267.9390^{\circ}$ \\
        %\hline
    \end{tabular}
\end{table}
%%%%%% Navid'version
%\subsection{Learning Setup}%%% Change it 
%%%% add two line 
\begin{figure}[t]
    \centering
    \includegraphics[width=0.42\textwidth]{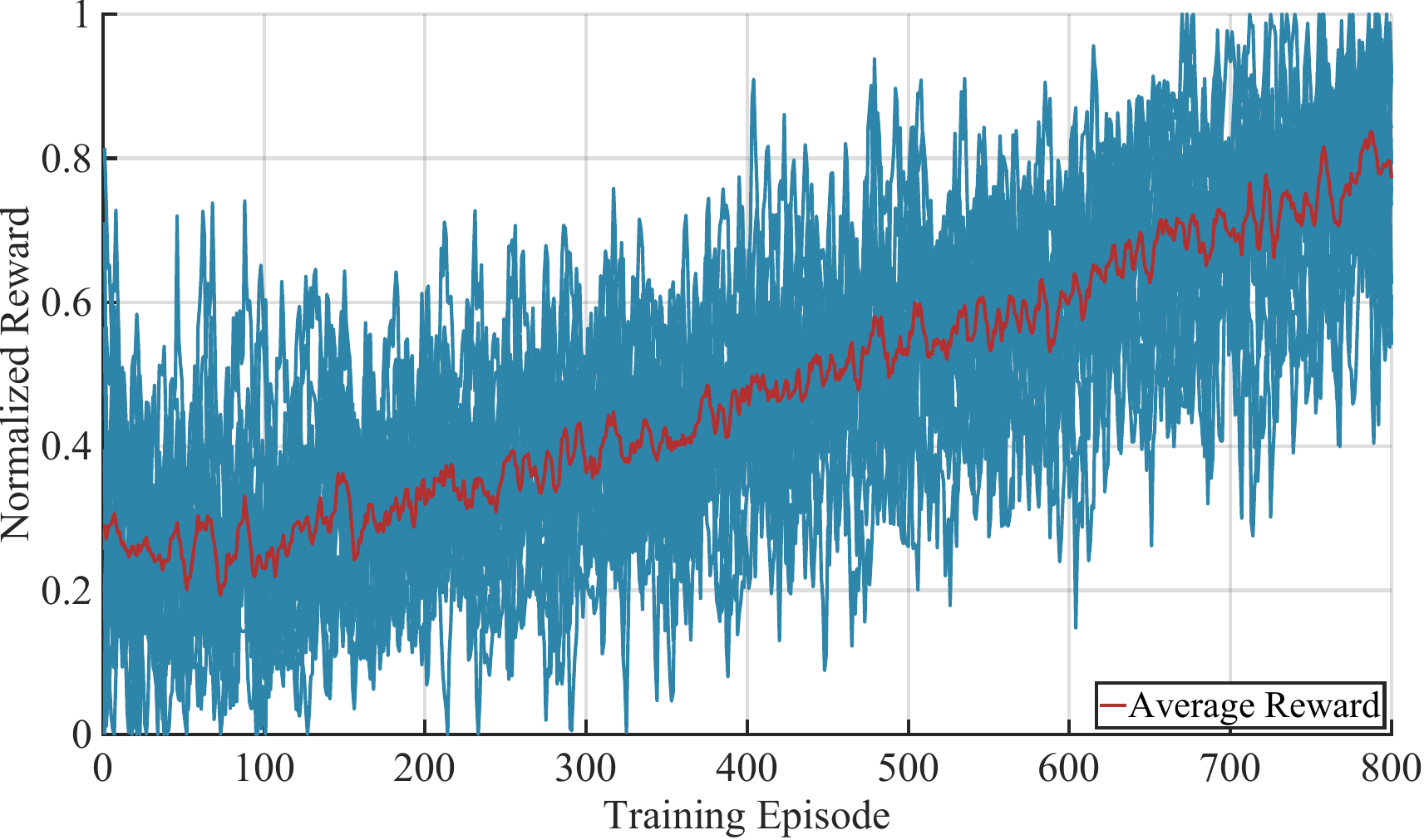}
    \caption{PPO Training Convergence across Random Seeds}
    \label{fig:PPO Training Convergence across Random Seeds}
\end{figure}
\subsection{Results and discussion}
\textbf{\textit{PPO Training Convergence}:} Figure \ref{fig:PPO Training Convergence across Random Seeds} illustrates the PPO training convergence across random seeds. Blue traces represent the normalized episodic rewards, while the red curve denotes their running average. The steadily increasing average reward indicates stable learning and consistent policy improvement, despite inherent variability in individual episodes typical of stochastic interference-management environments.
%Each blue trace denotes the normalized reward obtained in an episode, while the red curve is the running average. The average reward exhibits a clear, monotonic upward trend over the training horizon, demonstrating that the learner consistently improves the policy with experience. Although individual episode rewards remain variable, expected in a stochastic interference-management setting, the increasing average and the gradual uplift in the reward envelope indicate robust learning progress and convergence toward higher objective values.
%%%%%% Navid'version
% Figure \ref{fig:PPO Training Convergence across Random Seeds} illustrates PPO training performance across random seeds. Each blue trace denotes the normalized reward obtained in an episode, while the red curve is the running average. The average reward exhibits a clear, monotonic upward trend over the training horizon, demonstrating that the learner consistently improves the policy with experience. Although individual episode rewards remain variable, expected in a stochastic interference-management setting, the increasing average and the gradual uplift in the reward envelope indicate robust learning progress and convergence toward higher objective values.

\begin{figure}[t]
    \centering
    \includegraphics[width=0.46\textwidth]{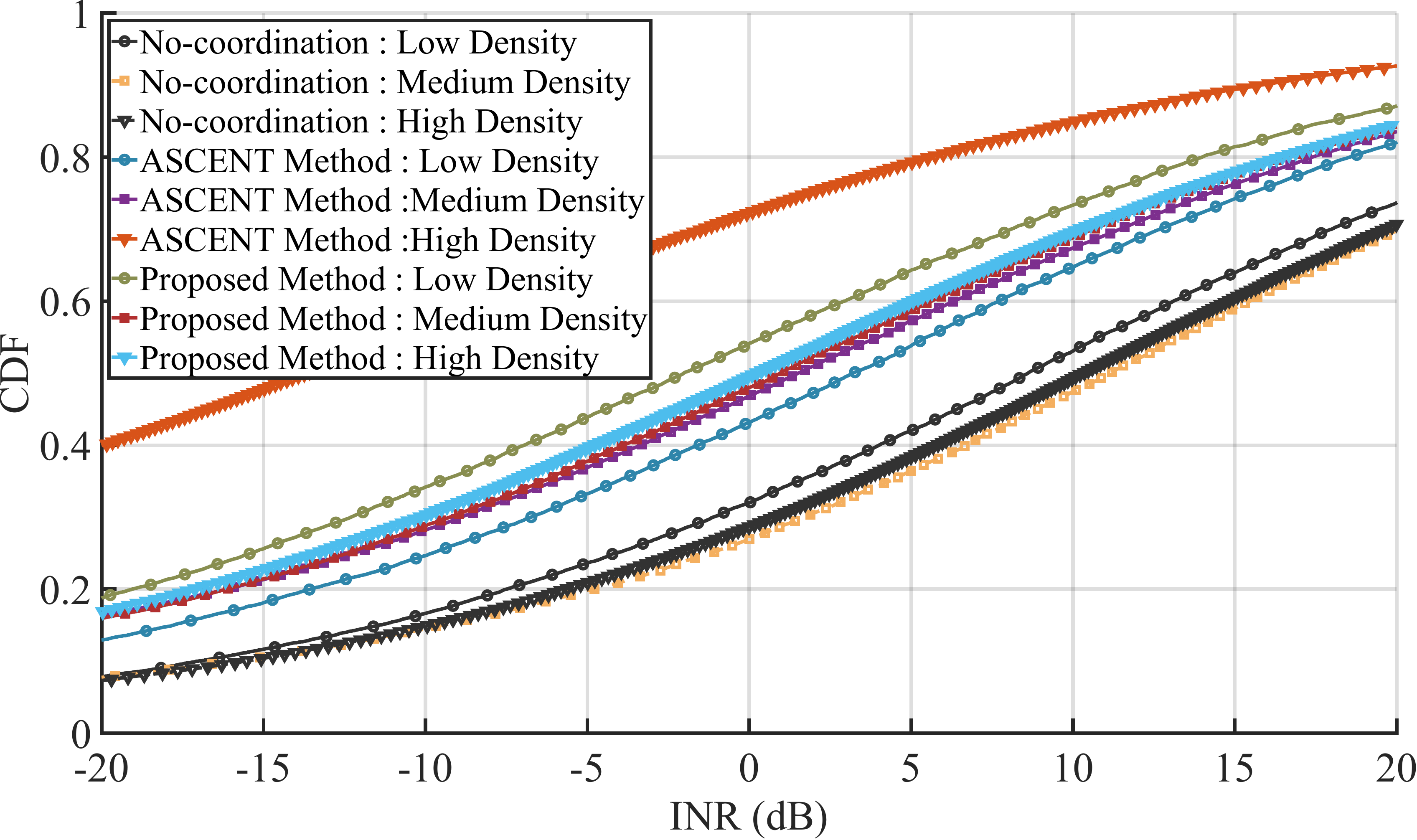}
    \caption{INR Comparison across NTN Densities}
    \label{fig:INR Comparison across NTN Densities}
\end{figure}

\textbf{\textit{Effect of NTN user density}:} Figure~\ref{fig:INR Comparison across NTN Densities} compares INR cumulative distribution functions for low (\(6\times10^{-8}\)), medium (\(1\times10^{-7}\)), and high (\(3\times10^{-7}\)) NTN user densities under different interference-management schemes. Here, only T1 fixed gateway NTN users are considered. Across all densities, the proposed PPO-based method provides robust interference suppression compared to both ASCENT and the no-coordination baseline. While ASCENT achieves its best performance under high-density conditions, where its threshold-based muting is frequently triggered, but its static exclusion rules become overly conservative and inefficient at lower densities. In contrast, the proposed learning-based framework adapts its actions to the INR threshold, leveraging selective sector muting, transmit-power control, and antenna downtilt adjustments. This enables efficient interference mitigation without resorting to large-scale TN shutdowns, thereby maintaining stable NTN protection across all densities.
%Figures~\ref{fig:INR Comparison across NTN Densities} shows our proposed DRL method yields a consistent left shift relative to the no-coordination baseline at low, medium, and high densities, showing robust interference reduction. ASCENT is most effective in the high-density regime; its threshold rule triggers frequent muting and produces the lowest INR there. At low density, however, ASCENT underperforms DRL which is reflecting the rigidity of rule-based muting when few NTN users are present. Overall, our proposed DRL delivers stable gains across densities without resorting to blanket shutdowns.

% \begin{figure}[t]
%     \centering
%     \includegraphics[width=0.45\textwidth]{Figs/INR_Comparison_across_NTN_Densities.pdf}
%     \caption{INR Comparison across NTN Densities}
%     \label{fig:INR Comparison across NTN Densities}
% \end{figure}
\begin{figure}[t]
    \centering
    \includegraphics[width=0.46\textwidth]{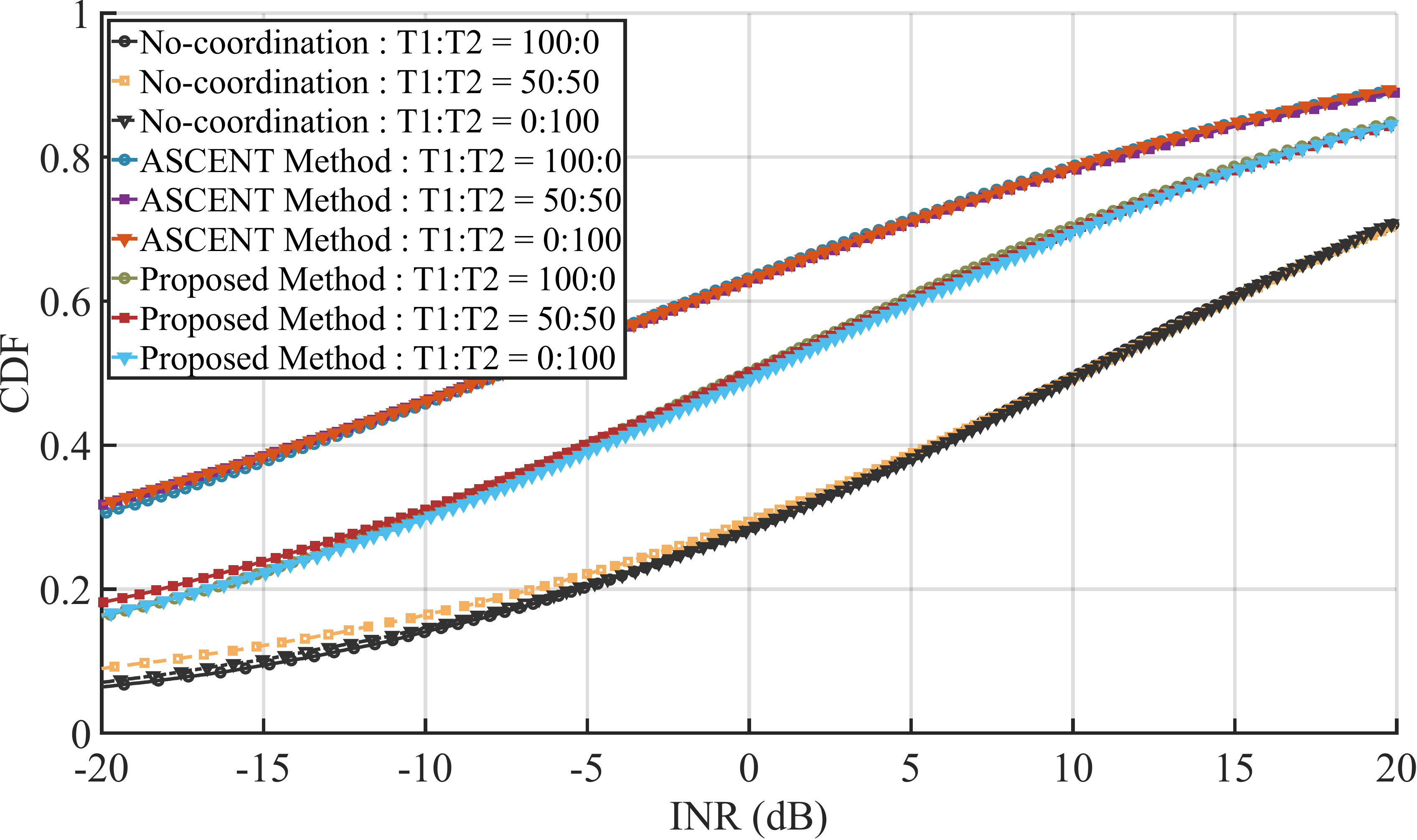}
    \caption{INR Comparison across NTN User Types}
    \label{fig:INR Comparison across NTN User Types}
\end{figure}

\textbf{\textit{Effect of NTN User Types}:} Figure \ref{fig:INR Comparison across NTN User Types} compares INR distributions for varying T1:T2 NTN user ratios, where $100:0$ (all fixed gateways), $50:50$ (mixed), and $0:100$ (all direct-to-cell devices). Scenarios dominated by T1 users yield lower INR due to their high-gain, narrow-beam antennas, while higher proportions of T2 users result in greater interference exposure. Across all user mixes, the proposed PPO-based approach achieves median INR improvements of $6$–$8$ dB over ASCENT and the no-coordination baseline. The proposed framework effectively suppresses TN interference for both user types, with clear gains for direct-to-cell users.

\begin{figure}[htbp]
    \centering
    \includegraphics[width=0.4\textwidth]{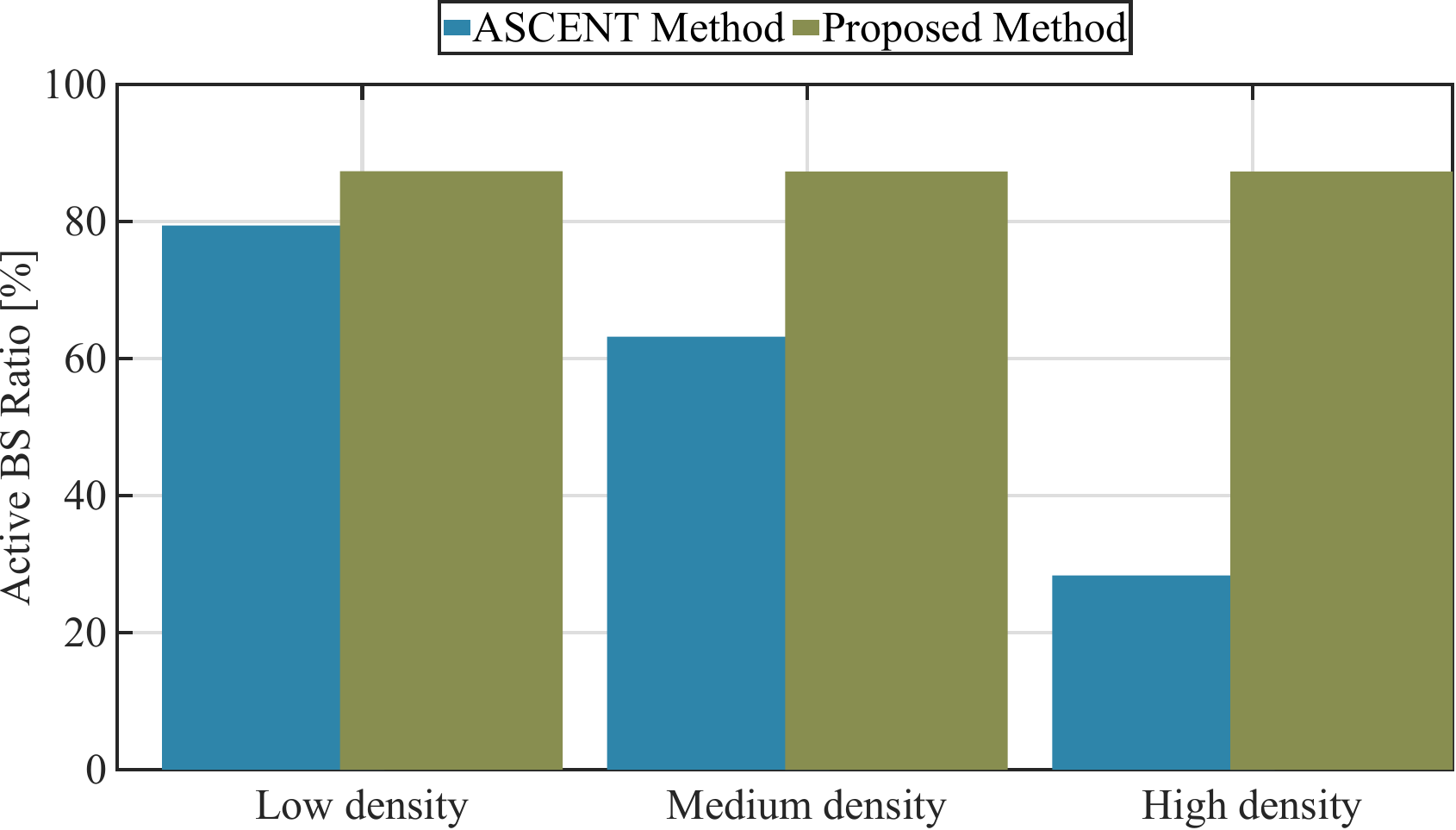}
    \caption{TN Base Station Activeness Comparison}
    \label{fig:TN Base Station Activeness Comparison}
\end{figure}

\textbf{\textit{Effect on TN Base-Station Activeness}: }Figure \ref{fig:TN Base Station Activeness Comparison} illustrates the active BS ratio across NTN densities, reflecting how each scheme balances NTN protection with TN service continuity. ASCENT shows a steep decline in activeness, from \(79.43\%\) at low density to \(28.37\%\) at high density, indicating that its lower INR (as shown in Figure~\ref{fig:INR Comparison across NTN Densities}) comes primarily from extensive muting rather than efficient interference management. In contrast, the proposed RL framework sustains a consistently high activeness level (\(\sim 87\%\)) across densities, supported by its adaptive per-sector control. This demonstrates that the proposed approach not only achieves the interference suppression trends observed in Fig. \ref{fig:INR Comparison across NTN Densities} and Fig. \ref{fig:INR Comparison across NTN User Types} but does so while preserving TN operational availability and avoiding unnecessary muting
%shows how each approach affects terrestrial network service using the base-station (BS) activeness ratio (lower activeness indicates more aggressive shutdown). The rule-based ASCENT method exhibits a steep decline with density, dropping from \(79.4299\%\) at low density to \(28.3742\%\) at high density, confirming that its lowest INR at high density is achieved via extensive muting and the attendant loss of TN service. By contrast, the proposed RL strategy maintains a consistently high activeness level, keeping \(\sim 87\%\) of BSs operational across all densities through selective sector activation. Overall, RL attains robust, low-INR performance with significantly better resource efficiency and service continuity than the non-adaptive ASCENT rule.

% \begin{figure}[t]
%     \centering
%     \includegraphics[width=0.42\textwidth]{Figs/INR_Comparison_across_NTN_User_Types.pdf}
%     \caption{INR Comparison across NTN User Types}
%     \label{fig:INR Comparison across NTN User Types}
% \end{figure}

\section{Conclusion}\label{sec:conclusion}
In this work, we developed an RL–based interference management framework to enable practical TN–NTN coexistence in the FR3 band. By jointly optimizing TN downlink power, antenna downtilt, and selective muting through a centralized PPO agent, the proposed approach dynamically adapts to varying interference and network conditions. Simulation results confirmed that the PPO-based control policy substantially enhances NTN downlink throughput and interference protection while maintaining TN activity in the overlap region, outperforming static power control and exclusion-zone methods. 
%In future, we plan to extend this framework toward fully distributed multi-agent learning architectures to reduce centralization overhead and enhance scalability in large TN deployments.
\section*{Acknowledgment}
\small
\vspace{-0.06cm}
This work was partially funded by the Federal Ministry of Education and Research Germany within the project “Open6GHub” under grant 16KISK012.
\bibliographystyle{IEEEtran}
\bibliography{TN_NTN.bib}

\end{document}